\documentclass[sigconf]{acmart}

\setcopyright{none}

\acmConference[Preprint]{}{May 2026}{Working Paper}
\acmISBN{}
\acmDOI{}
\renewcommand\footnotetextcopyrightpermission[1]{}
\usepackage{graphicx}

\usepackage{amsmath,amssymb}
\usepackage{booktabs}
\usepackage{multirow}
\usepackage{algorithm}
\usepackage{algorithmic}
\usepackage{listings}
\usepackage[most]{tcolorbox}
\usepackage{xcolor}

\lstdefinestyle{myCode}{
    language=, 
    basicstyle=\ttfamily\small,
    breaklines=true,
    showstringspaces=false,
    frame=none,
    backgroundcolor=, 
    aboveskip=0pt,
    belowskip=0pt,
    xleftmargin=0pt
}

\newtcolorbox{examplebox}[2][]{
  enhanced,
  colback=gray!5, 
  colframe=gray!60, 
  fonttitle=\bfseries, 
  coltitle=white,      
  colbacktitle=gray!70, 
  title=#2, 
  attach boxed title to top left={xshift=2mm, yshift=-2mm},
  boxed title style={arc=2mm, boxrule=0pt},
  arc=2mm, 
  boxrule=1pt,
  #1
}

\title{FinCom: A Financial Multi-Agent Demo with Disagree-or-Commit Deliberation}

\author{Chao Peter Yang}
\authornote{These authors contributed equally as co-first authors.}
\affiliation{%
  \institution{Duke University}
  \country{United States}
}
\email{peter.yang@duke.edu}

\author{Zixiao Tan}
\authornotemark[1]
\affiliation{%
  \institution{Duke University}
  \country{United States}
}
\email{zixiao.tan@duke.edu}

\author{Kaisen Yao}
\authornotemark[1]
\affiliation{%
  \institution{Duke University}
  \country{United States}
}
\email{kaisen.yao@duke.edu}

\author{Ziyu Zhou}
\authornote{These authors contributed equally as co-second authors.}
\affiliation{%
  \institution{Duke University}
  \country{United States}
}
\email{ziyu.zhou2@duke.edu}

\author{Eleanor Jiang}
\authornotemark[2]
\affiliation{%
  \institution{Duke University}
  \country{United States}
}
\email{eleanor.jiang@duke.edu}

\author{Michael Wu}
\affiliation{%
  \institution{ClearPath}
  \country{United States}
}
\email{michael@clearpathagent.com}

\begin{teaserfigure}
  \centering
  \includegraphics[width=\textwidth]{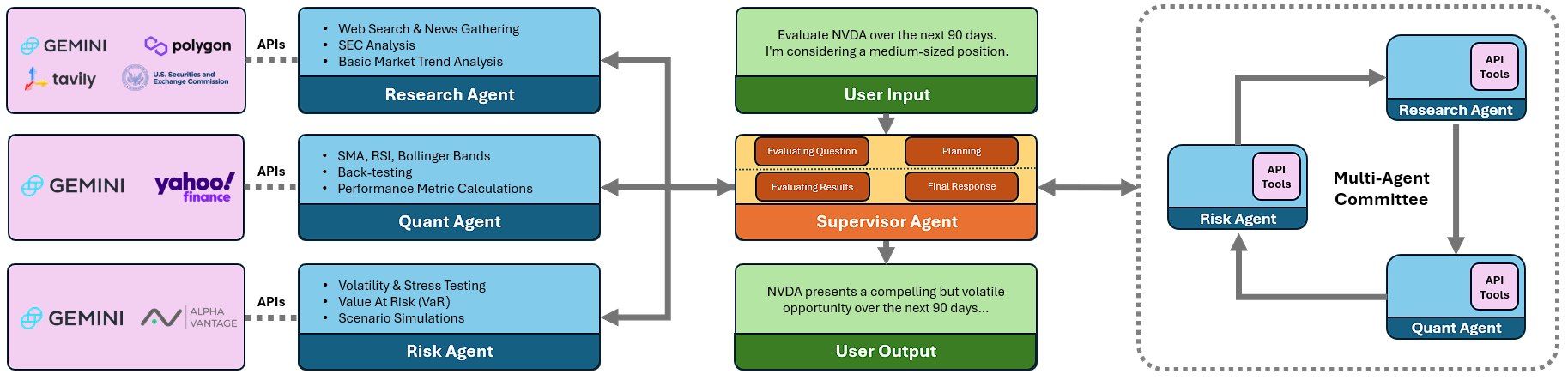}
  \caption{\textbf{FinCom architecture with Disagree-or-Commit (DoC).} A Supervisor coordinates three specialist agents---Research, Quant, and Risk Management---for financial analysis. In committee mode, each agent reviews prior reasoning and must either critique it or commit to it before the Supervisor synthesizes the final response.}
  \Description{A system diagram showing a supervisor agent coordinating three specialized agents: Research, Quantitative, and Risk. Each agent performs independent analysis before entering a structured Disagree-or-Commit phase that forces explicit dissent before final consensus and decision output.}
  \label{fig:architecture}
\end{teaserfigure}

\begin{document}

\begin{abstract}
Multi-agent systems powered by large language models (LLMs) are increasingly used for financial analysis and decision support. However, existing coordination schemes, especially those emphasizing consensus or debate, are vulnerable to \textit{sycophancy}: agents conform to peer reasoning instead of evidence, leading to premature agreement and degraded outcomes. We introduce \textbf{FinCom} (Financial Committee), a governed multi-agent framework and interactive system that operationalizes the \textbf{Disagree-or-Commit (DoC)} protocol to embed structured dissent into financial AI committees. A central \textbf{Supervisor} orchestrates three ReAct-enabled specialist agents---\textbf{Research}, \textbf{Quantitative}, and \textbf{Risk}---each equipped with role-specific tools for retrieval, computation, and stress testing. During deliberation, agents must either explicitly critique or commit to their peers’ reasoning before converging on a unified recommendation. 

This demonstration showcases how FinCom supports committee-style financial analysis through coordinated multi-agent interaction, including structured report generation and interactive decision support. Evaluated across the most recent financial agent benchmark, in addition to 90 internal handcrafted financial tasks using an LLM-as-a-Judge protocol, DoC improves reasoning accuracy and risk awareness significantly over a consensus-seeking baseline on both an in-house and external evaluation set. By reframing disagreement as a governance primitive rather than noise, FinCom offers a lightweight, prompt-only recipe for improving accountability, transparency, and epistemic robustness in agentic financial systems.
\end{abstract}

\ccsdesc[500]{Computing methodologies~Artificial intelligence}
\ccsdesc[500]{Computing methodologies~Multi-agent systems}

\keywords{LLM agents, multi-agent systems, finance, disagreement, governance, sycophancy}

\maketitle

\section{Introduction}

Financial decision-making inherently requires combining qualitative research, quantitative analysis, and risk review---perspectives that are difficult to unify within a single LLM agent. Multi-agent systems offer a natural solution, but introduce a coordination problem: agents tend to over-align with earlier reasoning rather than independently validating evidence, producing outputs that appear coherent without being sufficiently grounded.

We present \textbf{FinCom}, an interactive financial multi-agent system that emulates a compact investment committee. To address conformity bias, FinCom adopts the \textbf{Disagree-or-Commit (DoC)} protocol, a prompt-layer coordination rule requiring each agent to either explicitly critique prior reasoning or endorse it with new supporting evidence. This makes deliberation auditable and discourages passive agreement.

The remainder of this paper describes the system architecture, the DoC protocol, demonstration scenarios, and supporting evaluation.

\section{Related Works}

\subsection{LLM-Based Financial Agents}
Financial LLM research has progressed from domain-adapted single agents---\textit{FinGPT} \citep{Yang2023}, \textit{PIXIU} \citep{Xie2023}---to role-specialized multi-agent systems such as \textit{TradingAgents} \citep{Xiao2025} and \textit{QuantAgent} \citep{Xiong2025}. Memory-augmented models like \textit{FinMem} \citep{Yu2023} further improved temporal reasoning. However, most frameworks remain consensus-oriented or rely on rigid debate structures, leaving conformity bias insufficiently addressed \citep{Xiao2025}.

\subsection{Coordinated Multi-Agent Systems}
The ReAct paradigm \citep{Yao2022} grounds agent reasoning through iterative tool use, while frameworks such as \textit{LangGraph} \citep{LangGraph2024} and \textit{MetaGPT} \citep{Hong2024} systematize multi-agent collaboration via graph-based routing and role orchestration. These approaches prioritize efficiency and modularity but seldom address the social dynamics of collective deliberation---overconformity, persuasion cascades, or the dominance of confident but inaccurate agents. FinCom's DoC protocol targets precisely this gap.

\section{System Overview}

FinCom is an interactive financial multi-agent system designed to emulate a compact investment committee. The system combines role-specialized agents, tool-augmented reasoning, and a governed coordination layer into a single end-to-end application. At a high level, FinCom receives a user request, routes it through a \textbf{Supervisor}, invokes one or more specialist agents, and returns either a structured report or an interactive committee response.

\subsection{Design Goals}
The system is built around three design goals:

\begin{itemize}
    \item \textbf{Role specialization}: separate qualitative research, quantitative analysis, and risk review into distinct agent roles.
    \item \textbf{Interactive usability}: support both one-shot report generation and conversational interaction.
    \item \textbf{Governed deliberation}: encourage specialists to validate and challenge one another rather than merely concatenate outputs.
\end{itemize}

\subsection{Architecture}
As shown in Figure \ref{fig:architecture}, FinCom consists of a central \textbf{Supervisor} and three specialist agents: \textbf{Research}, \textbf{Quant}, and \textbf{Risk Management}. Each agent is implemented as a LangGraph node with a role-specific system prompt and constrained tool access. All agents operate in a prompt-only configuration without task-specific fine-tuning, which keeps the system modular and easy to adapt to new financial workflows.

The Supervisor is responsible for intent parsing, task decomposition, routing, and final synthesis. Depending on the request, it can either call a single specialist directly or convene the full committee. Direct invocation is useful for narrow tasks such as computing a technical indicator or retrieving filing-based evidence. Committee invocation is used for higher-level investment questions that benefit from multiple perspectives.

\subsection{Execution Flow}
Given a user query, the Supervisor first determines whether the request is best handled by a single agent or by committee deliberation. It then generates a structured sub-task and dispatches it to the relevant agent or agents.

In single-agent mode, the selected specialist executes a ReAct loop (reason $\rightarrow$ act $\rightarrow$ observe) using its role-specific tools and returns a grounded answer. In committee mode, the Supervisor sequentially invokes the specialists, passing forward the accumulated reasoning trace. Each subsequent agent reviews the prior analysis, contributes its own perspective, and may call additional tools before returning its response to the Supervisor. The Supervisor then synthesizes the committee output into a final user-facing answer.

\subsection{Research Agent}
The Research Agent performs qualitative evidence gathering and synthesis across five tool-augmented capabilities:
\begin{itemize}
  \item \textbf{Earnings intelligence}: retrieves earnings call transcripts with speaker-level attribution via the DefeatBeta API.
  \item \textbf{Fundamental data}: queries financial ratios, analyst ratings, price targets, insider transactions, and social sentiment via Finnhub.
  \item \textbf{SEC filings}: extracts and AI-summarizes 10-K, 10-Q, and 8-K reports, including MD\&A sections and risk factors.
  \item \textbf{Technical indicators}: computes SMA, EMA, RSI, and MACD over historical OHLC data via the Polygon API.
  \item \textbf{Web search}: retrieves macro trends, industry analysis, and breaking news via Tavily.
\end{itemize}


\subsection{Quant Agent}
The Quant Agent performs technical and statistical analysis over historical market data. It produces reproducible, structured evidence to support investment reasoning. Its main functions include:
\begin{itemize}
  \item \textbf{Technical indicators}: computing moving averages, RSI, MACD, Bollinger Bands, and related signals.
  \item \textbf{Backtesting}: evaluating simple rule-based strategies over historical data and summarizing return and risk metrics.
  \item \textbf{Cross-asset analysis}: examining volatility, beta, and correlation structure relevant to portfolio construction.
  \item \textbf{Visualization}: producing interpretable charts that support downstream explanation and review.
\end{itemize}

Representative indicators and metrics include:
\begin{align}
\mathrm{SMA}_t^{(n)} &= \frac{1}{n}\sum_{i=0}^{n-1} C_{t-i}, \nonumber\ \mathrm{MACD}_t = \mathrm{EMA}_t^{(12)} - \mathrm{EMA}_t^{(26)}, \nonumber\\
\mathrm{RSI}_t &= 100 - \frac{100}{1 + RS_t}, \quad
RS_t = \frac{\mathrm{AvgGain}_t}{\mathrm{AvgLoss}_t}, \nonumber\\
R_t &= \frac{C_t - C_{t-1}}{C_{t-1}}, \quad
\rho_{ij} = \frac{\mathrm{Cov}(R_i, R_j)}{\sigma_i \sigma_j}. \nonumber
\end{align}

\subsection{Risk Management Agent}
The Risk Management Agent translates quantitative evidence into downside-aware portfolio assessment. It focuses on exposure, drawdown, and scenario sensitivity, complementing the more opportunity-focused analyses of the other agents. Its main functions include:
\begin{itemize}
  \item \textbf{Volatility and drawdown analysis}: estimating realized risk and historical downside behavior.
  \item \textbf{Value-at-Risk (VaR)}: estimating tail loss under historical simulation.
  \item \textbf{Stress testing}: probing portfolio behavior under adverse or concentrated scenarios.
  \item \textbf{Qualitative risk review}: surfacing non-price risks such as regulatory, geopolitical, and concentration concerns.
\end{itemize}

Representative risk metrics include:
\begin{align}
r_t &= \ln(P_t/P_{t-1}), \nonumber\\
\sigma_{\text{annual}} &= \sigma(r_t)\sqrt{252}, \nonumber\\
\text{MaxDD} &= \min_t \left( \frac{P_t - \max_{i \le t} P_i}{\max_{i \le t} P_i} \right), \nonumber\\
\text{VaR}_1 &= P_0 \cdot |\text{percentile}_{1-\alpha}(r_t)|. \nonumber
\end{align}

\subsection{Disagree-or-Commit (DoC) Protocol}
FinCom uses a lightweight coordination rule called \textbf{Disagree-or-Commit (DoC)} during committee deliberation. Before producing its response, each agent must review the prior agent’s reasoning and take one of two actions:
\begin{enumerate}
    \item \textbf{Disagree}: explicitly identify an error, contradiction, unsupported claim, or missing piece of evidence, and provide corrective reasoning or evidence; or
    \item \textbf{Commit}: explicitly endorse the prior reasoning and add at least one new supporting fact, clarification, or extension.
\end{enumerate}

This protocol serves two purposes. First, it prevents committee interaction from degenerating into passive agreement. Second, it yields a more auditable reasoning trace by making agreement and disagreement explicit. Unlike heavier debate-style frameworks \citep{irving2018ai, du2023debate, smit2024mad}, DoC is implemented purely at the prompt layer and can be added to an existing supervisor-based architecture without retraining.

\section{Demonstration Scenarios}

\begin{figure}
    \centering
    \includegraphics[width=1.0\linewidth]{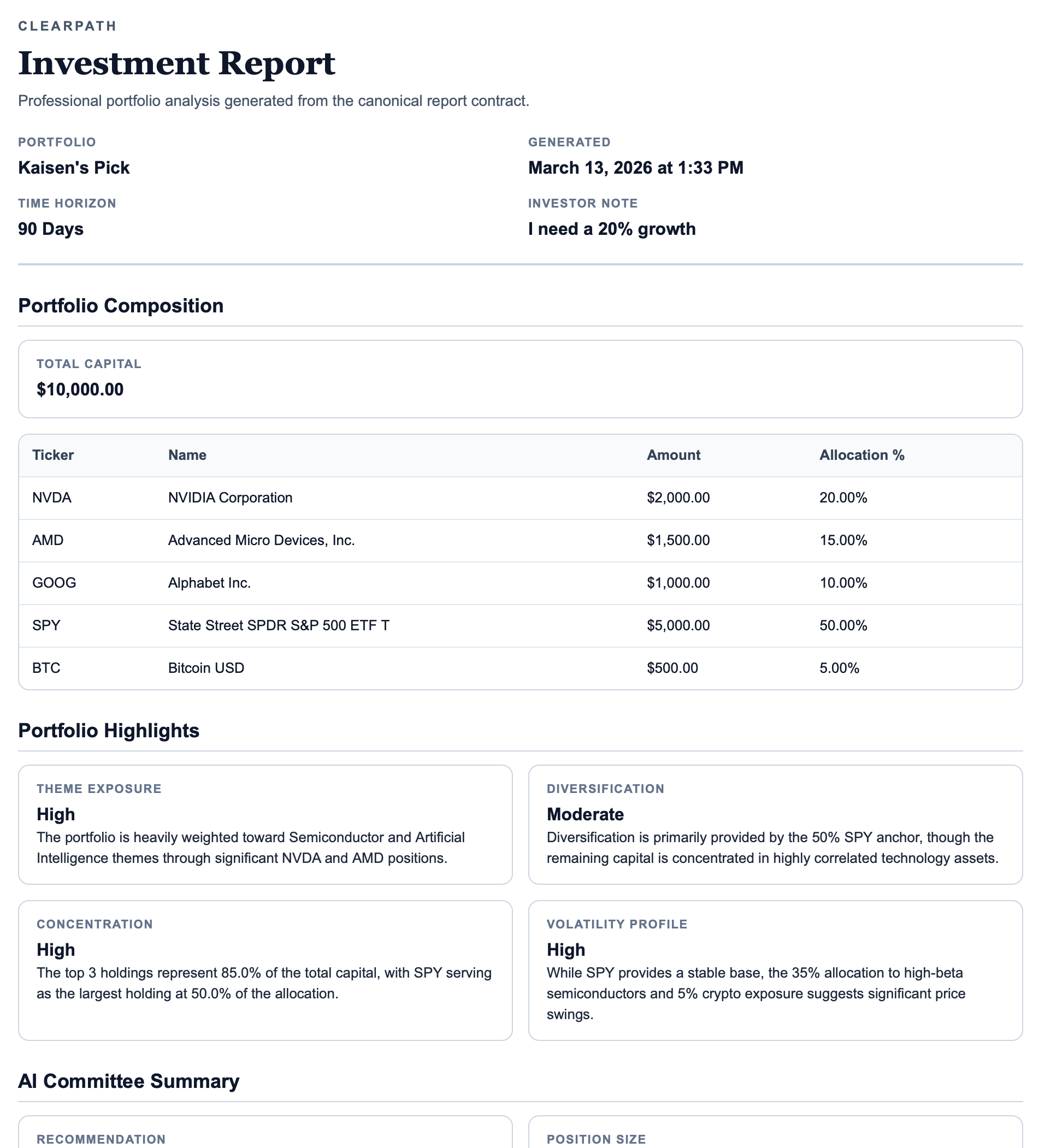}
    \caption{Sample Report as Generated by FinCom}
    \label{fig:placeholder}
\end{figure}

The core contribution of this paper is a working demonstration of FinCom as an interactive financial AI system. The demo is organized around two user-facing scenarios that showcase how the underlying multi-agent architecture supports realistic analyst workflows.

\subsection{Scenario 1: Report Generation}
FinCom can generate structured reports for either a single security or an entire portfolio. In this workflow, the user submits a stock or portfolio, and the Supervisor determines whether to invoke a single specialist or convene the full committee. The returned report is designed to resemble an internal investment memo, with separate sections corresponding to the committee’s perspectives.

A typical generated report includes:
\begin{itemize}
    \item \textbf{Committee Summary}: a concise synthesis of the overall recommendation and key takeaways;
    \item \textbf{Research View}: recent company developments, macro context, filings-based evidence, and relevant news;
    \item \textbf{Quantitative View}: technical indicators, historical trends, backtest summaries, and supporting figures;
    \item \textbf{Risk View}: concentration concerns, volatility and drawdown statistics, and scenario-based risk commentary;
    \item \textbf{References}: cited external sources used by the system;
    \item \textbf{Decision Signals}: summary indicators such as stance, confidence, and major watch-outs.
\end{itemize}

This workflow highlights FinCom’s ability to transform a user query into a transparent, multi-perspective financial report rather than a single undifferentiated response.

\subsection{Scenario 2: Committee Chat}

In the second scenario, users interact with FinCom through a committee-chat interface. They may consult a single specialist directly or engage the full committee in a shared discussion. This supports prompts such as analyzing a portfolio, comparing securities, requesting diversification suggestions, or asking the system to justify a recommendation from a research-first, quant-first, or risk-first perspective.

A key advantage of this mode is that users can observe how specialist viewpoints are surfaced and reconciled. Rather than hiding deliberation behind a final answer, FinCom exposes the structure of the interaction and makes disagreement or endorsement between agents more legible.

\subsection{What the Live Demo Shows}
In a live demonstration, we show the full end-to-end pipeline: user query intake, Supervisor routing, specialist tool use, committee deliberation, and final response generation. We also highlight how DoC changes the interaction trace by forcing explicit endorsement or critique between agents. This makes the demo not just a showcase of financial question answering, but of governed multi-agent coordination in a realistic application domain.

We do not provide a public live demo URL because the current prototype requires substantial deployment, orchestration, and maintenance effort to support reliable external access, which is beyond the scope of our academic research setting. Instead, the system is demonstrated through an end-to-end video showing the complete workflow, intermediate reasoning traces, and final outputs.

\section{Evaluation}

Although this paper is centered on the system demonstration, we include an empirical evaluation to assess whether the committee architecture and DoC protocol improve the quality of FinCom’s outputs.

\subsection{Datasets}
We evaluate FinCom on two complementary data sources.

First, we use the public validation split of a financial agent benchmark (refereed to as FinAgent Bench hereafter) containing 50 open-source examples spanning retrieval, reasoning, forecasting, and market analysis by Bigeard et al \cite{bigeard2025fab}. The benchmark targets capabilities expected of an entry-level financial analyst and provides reference answers for evaluation \cite{bigeard2025fab}.

Second, we construct an internal dataset of 90 evaluation instances tailored to multi-agent financial workflows. These are divided into three categories with 30 examples each: \textbf{research-heavy}, \textbf{quant-heavy}, and \textbf{risk-focused} tasks. Unlike the benchmark subtasks, these instances were designed to reflect end-to-end committee-style reasoning and the kinds of interactions showcased in the live system.

\subsection{Protocol}
As illustrated in Figure~\ref{fig:eval} of the Appendix, we use an \textit{LLM-as-a-Judge} pipeline to compare system outputs against reference answers. Each response is graded using a task-specific evaluation prompt. A score of 1 is assigned when the output contains the required information without contradiction, and 0 otherwise. To reduce stochasticity, each evaluation is repeated and average scores are reported.

\section{Results}\label{sec:res}

Table~\ref{tab:dc-ablation} shows that the benefits of FinCom are most consistent on the internal financial workflows, where committee-based coordination and DoC improve performance over simpler baselines.

On the internal tasks, \textbf{Supervisor-Committee + DoC} achieves the best results across all three categories: \textbf{54.2\%} on Research, \textbf{96.3\%} on Quant Finance, and \textbf{90.5\%} on Risk Management. The largest gain appears in \textbf{Risk Management}, where DoC improves substantially over standard committee coordination (90.5\% vs.\ 59.3\%), suggesting that explicit critique is especially helpful for downside-sensitive analysis. Quantitative tasks remain a relative strength across multi-agent variants, with DoC further improving over \textbf{Supervisor + Agents} (96.3\% vs.\ 92.9\%).

On the external \textbf{FinAgent} benchmark, the pattern is more mixed. \textbf{Supervisor-Committee} achieves the highest score at \textbf{66.0\%}, while \textbf{Supervisor-Committee + DoC} reaches 58.7\%. This suggests that DoC is most beneficial for longer, workflow-oriented tasks requiring multi-perspective reconciliation, rather than shorter benchmark-style questions.

Overall, these results support the main claim of the demo: FinCom is not just a collection of specialist agents, but a governed coordination framework. The combination of role specialization, committee interaction, and explicit disagreement handling produces the strongest results on realistic financial analysis tasks.

\section{Discussion}\label{sec:lim}

FinCom demonstrates how role-specialized agents and lightweight governed deliberation can be combined in a usable financial AI system. While our evaluation provides supporting evidence that DoC improves performance on realistic internal workflows, broader human evaluation and user studies are needed to assess trust, usability, and generalization. Future work includes comparing DoC with other coordination mechanisms and extending FinCom with additional specialist agents and richer portfolio constraints.

\section{Conclusion}\label{sec:con}

We presented \textbf{FinCom}, an interactive financial multi-agent system built to support committee-style investment analysis. FinCom combines a Supervisor-led architecture, role-specialized agents for research, quantitative analysis, and risk management, and a lightweight \textbf{Disagree-or-Commit} protocol that makes inter-agent validation explicit.

As a demonstration system, FinCom contributes an end-to-end example of how governed multi-agent coordination can be translated into a usable financial application. The system supports both structured report generation and interactive committee chat, allowing users to observe how different specialist perspectives are produced and reconciled. Our evaluation provides supporting evidence that this coordination design improves response quality on realistic financial tasks.

Taken together, FinCom shows that practical financial AI systems can benefit from treating deliberation as a first-class systems problem. By combining specialization, tool grounding, and explicit coordination, the demo illustrates a path toward more transparent and auditable agentic decision support.

\section{Acknowledgment and LLM Disclosure}
The system design, methodology, experiments, and analysis were developed by the authors. LLM assistance was used only for language polishing and editorial refinement of the manuscript.

We also thank Mimi, the authors' cat, for steadfast moral support during the preparation of this paper.

\label{sec:ref}
\bibliographystyle{ACM-Reference-Format}
\bibliography{bibliography}

\appendix
\section{Appendix}\label{sec:app}
\begin{figure*}
    \centering
    \includegraphics[width=\linewidth]{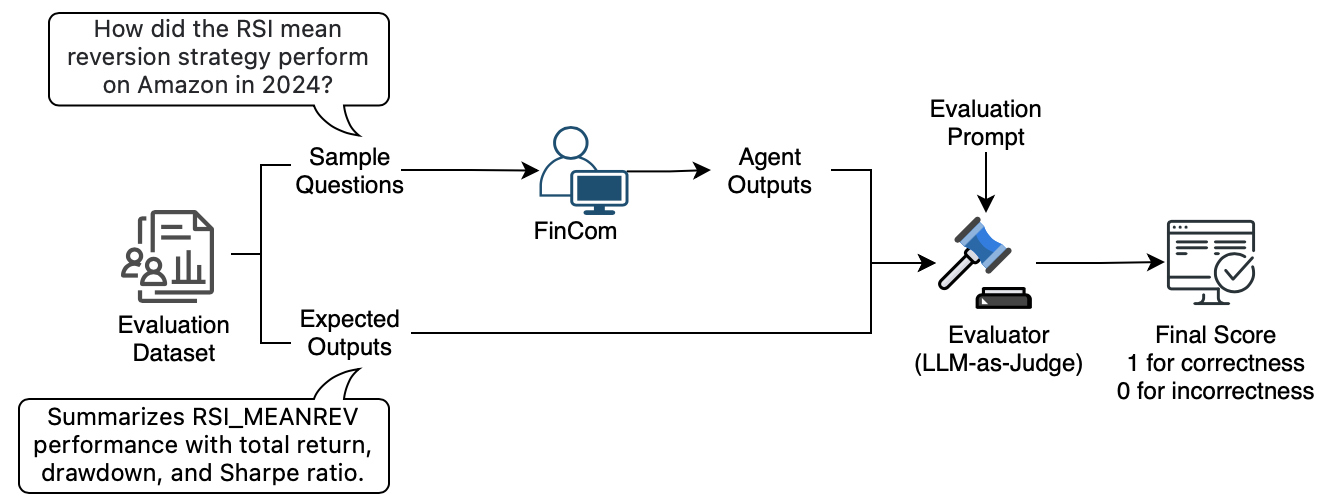}
    \caption{\textbf{LLM-as-a-Judge Evaluation Pipeline}: The output of the FinCom agent is scored compared to the reference output for each data point by the Judge LLM agent, which, in this case, is using Gemini-2.5-Flash}
    \label{fig:eval}
\end{figure*}

This appendix provides supplementary evaluation and implementation details omitted from the main paper for space reasons. In particular, we include the LLM-as-a-Judge evaluation pipeline (Figure~\ref{fig:eval}), the full quantitative results table (Table~\ref{tab:dc-ablation}), the Disagree-or-Commit prompt used in committee deliberation, the evaluator prompts, and representative examples from our LangSmith experiments.

\begin{table*}[]
\centering
\begin{tabular}{lccccc}
\hline
\textbf{Method} & \textbf{Research} & \textbf{Quant Finance} & \textbf{Risk Management} & \textbf{FinAgent} & \textbf{Overall Avg} \\
\hline

LLM & 53.3\% & 66.7\% & 73.3\% & 24.0\% & 54.3\% \\

LLM + Tools (ReAct) & 36.7\% & 73.3\% & 53.3\% & 46.0\% & 52.3\% \\

Supervisor + Agents & 42.9\% & 92.9\% & 70.8\% & 52.0\% & 64.7\% \\

Supervisor-Committee & 46.4\% & 88.9\% & 59.3\% & \textbf{66.0}\% & 65.2\% \\

\textbf{Supervisor-Committee + DoC} & \textbf{54.2}\% & \textbf{96.3}\% & \textbf{90.5}\% & 58.7\% & \textbf{74.9}\% \\

\hline
\end{tabular}
\caption{Performance comparison across financial reasoning tasks and the FinAgent benchmark for different agent architectures. Scores represent the proportion of correct or complete reasoning steps according to LLM-as-a-Judge evaluation. The underlying model is Gemini 3.1 Pro.}
\label{tab:dc-ablation}
\end{table*}

\begin{figure*}
  \begin{examplebox}{Disagree or Commit Prompt}
    \begin{lstlisting}[style=myCode, xleftmargin=2mm]
### Disagree or Commit (DoC)
Before producing your response, review reasoning from the previous agent.  
- If you detect errors, contradictions, or missing evidence, **state them clearly** and provide **corrective evidence** or reasoning with citations for the supervisor agent.   
- If the reasoning is valid, **acknowledge agreement** and add at least one new supporting fact or contextual clarification.  

Think step by step.
    \end{lstlisting}
  \end{examplebox}
    \caption{Disagree or Commit Prompt}
  \label{fig:doc_example}
\end{figure*}

\begin{figure*}[t!]
  \centering

  \begin{examplebox}{Evaluator System Prompt Base}
    \begin{lstlisting}[style=myCode, xleftmargin=2mm]
You are a correctness evaluation expert for a finance agent. If the output contains all the information of the reference output and doesn't contain contradictory statements, score 1, else 0. 
    \end{lstlisting}
  \end{examplebox}

  \begin{examplebox}{Evaluator Prompt Template}
    \begin{lstlisting}[style=myCode, xleftmargin=2mm]
Please grade the following agent run given the input, expected output, and actual output:

<example>
<input>
{{input}}
</input>

<output>
{{output}}
</output>

<referenceOutput>
{{referenceOutput}}
</referenceOutput>
</example>
    \end{lstlisting}
  \end{examplebox}
  \caption{Example Evaluation Prompt: (minor changes are made for each dataset)}
  \label{fig:eval_example}

\end{figure*}

\begin{figure*}[t!]
  \centering

  \begin{examplebox}{User Input (messages:)}
    \begin{lstlisting}[style=myCode, xleftmargin=2mm]
Find information about Federal Reserve interest rate decision in December 2024
    \end{lstlisting}
  \end{examplebox}

  \begin{examplebox}{Expected Output:}
    \begin{lstlisting}[style=myCode, xleftmargin=2mm]
The agent should use web_search with focused query on 'Federal Reserve interest rate decision December 2024'. Response should include meeting outcome, rate change details, Fed statement highlights, market reaction, and expert analysis from credible financial news sources.
    \end{lstlisting}
  \end{examplebox}
  
  \begin{examplebox}{FinCom Output:}
    \begin{lstlisting}[style=myCode, xleftmargin=2mm]
The Federal Reserve, through its Federal Open Market Committee (FOMC), decided on December 18, 2024, to lower the target range for the federal funds rate by 0.25 percentage point, bringing it to 4.25%-4.50%. This decision was communicated via an official press release and statement from the Federal Reserve.

Following the announcement, stocks experienced a sharp sell-off, and Treasury yields increased. Futures pricing also adjusted, showing a reduced expectation for further rate cuts in 2025.
    \end{lstlisting}
  \end{examplebox}
    \caption{Example of Research Dataset}
  \label{fig:backtest_example}

\end{figure*}

\begin{figure*}[t!]
  \centering
  \begin{examplebox}{User Input (messages:)}
    \begin{lstlisting}[style=myCode, xleftmargin=2mm]
Which strategy performed better for NVDA year 2023, SMA_CROSS or RSI_MEANREV?
    \end{lstlisting}
  \end{examplebox}

  \begin{examplebox}{Expected Output:}
    \begin{lstlisting}[style=myCode, xleftmargin=2mm]
Compares total return and Sharpe ratio of both backtests.
    \end{lstlisting}
  \end{examplebox}
  
  \begin{examplebox}{FinCom Output:}
    \begin{lstlisting}[style=myCode, xleftmargin=2mm]
For NVDA in 2023, the **RSI Mean Reversion (RSI_MEANREV)** strategy performed better than the **SMA Crossover (SMA_CROSS)** strategy.

Here's a breakdown of the performance:

**RSI Mean Reversion (RSI_MEANREV)**:
*   **Total Return**: -11.22% (a loss of $11,223 on a hypothetical $100,000 investment).
*   **Max Drawdown**: 18.97%.
*   **Annualized Volatility**: 19.71%.
*   **Sharpe Ratio (approx)**: -0.51.

**SMA Crossover (SMA_CROSS)**:
*   **Total Return**: -17.02% (a loss of $17,021 on a hypothetical $100,000 investment).
*   **Max Drawdown**: 45.22%.
*   **Annualized Volatility**: 49.22%.
*   **Sharpe Ratio (approx)**: -0.14.

**Comparison**:
While both strategies showed negative returns for NVDA in 2023, the RSI Mean Reversion strategy incurred a smaller loss (-11.22% vs. -17.02%). More significantly, it demonstrated considerably lower risk with a much smaller maximum drawdown (18.97% vs. 45.22%) and lower annualized volatility (19.71% vs. 49.22%). The Sharpe Ratio, while negative for both, was less negative for SMA_CROSS, which might seem counterintuitive given the other metrics. However, a less negative Sharpe Ratio with a significantly higher drawdown and volatility indicates that while its risk-adjusted return was slightly better, the overall risk taken was much higher, leading to a poorer absolute return and higher risk exposure. Therefore, based on overall return and risk metrics, RSI_MEANREV was the better performer, delivering a more controlled downside.
    \end{lstlisting}
  \end{examplebox}
    \caption{Example of quant agent executing a complex backtest query and providing a detailed performance and risk analysis.}
    \label{fig:backtest_example2}

\end{figure*}

\begin{figure*}[t!]
  \centering

  \begin{examplebox}{User Input (messages:)}
    \begin{lstlisting}[style=myCode, xleftmargin=2mm]
Is Johnson & Johnson (JNJ) considered a defensive stock?
    \end{lstlisting}
  \end{examplebox}

  \begin{examplebox}{Expected Output:}
    \begin{lstlisting}[style=myCode, xleftmargin=2mm]
Yes, JNJ is typically considered defensive. Its market beta is [X.XX], (usually < 1), indicating low volatility relative to the S&P 500. Its qualitative factors include stable demand for healthcare products, though it also faces litigation risks.
    \end{lstlisting}
  \end{examplebox}
  
  \begin{examplebox}{FinCom Output:}
    \begin{lstlisting}[style=myCode, xleftmargin=2mm]
    Yes, Johnson & Johnson (JNJ) is widely considered a defensive stock. This classification is supported by several factors:

*   **Outperformance during Market Stress:** JNJ has a history of outperforming the broader market (S&P 500) during periods of economic downturns and market volatility, such as the 2008 Financial Crisis, 2020 COVID Pandemic, and the 2022 Inflation Shock. This indicates its resilience in challenging economic environments.
*   **Price Stability:** The stock tends to exhibit price stability, reflecting consistent investor confidence even amidst broader economic headwinds.
*   **Diversified Healthcare Portfolio:** JNJ operates in essential healthcare sectors, including pharmaceuticals, medical devices, and consumer health. Products in these areas are generally in demand regardless of economic conditions, providing stable revenue streams.
*   **Strong Financial Health:** The company maintains a robust financial position with a strong cash flow and a conservative debt load, further contributing to its stability.
*   **Institutional Interest:** JNJ often attracts institutional investors seeking stability during volatile market periods.

In summary, JNJ's business model, financial strength, and historical performance align with the characteristics of a defensive stock, making it a relatively stable investment during economic uncertainties.
    \end{lstlisting}
  \end{examplebox}
\caption{Example of Risk Management Dataset}
  \label{fig:backtest_example3}

\end{figure*}

\end{document}